\documentstyle[aps]{revtex}   
\def\be{\begin{equation}}
\def\ee{\end{equation}}
\def\beq{\begin{eqnarray}}
\def\eeq{\end{eqnarray}}

\def\bay{\begin{array}}
\def\eay{\end{array}}

\begin{document}

\title{Spherically Symmetric, Self-Similar Spacetimes}

\author{Sanjay M. Wagh$^{\star}$
and  Keshlan S. Govinder$^{\dagger}$}

\address{$^{\star}$Central India Research Institute, Post Box 606,
Laxminagar, Nagpur 440 022, India. \\E-mail: ciri@vsnl.com}
\address{$^{\dagger}$School of Mathematical and Statistical
Sciences, University of Natal, Durban 4041, South Africa.
\\E-mail: govinder@nu.ac.za }
\date{December 14, 2001}
\maketitle

\begin{abstract}
Self-similar spacetimes are of importance to cosmology and to
gravitational collapse problems. We show that self-similarity or
the existence of a homothetic Killing vector field for spherically
symmetric spacetimes implies the separability of the spacetime
metric in terms of the co-moving coordinates and that the metric
is, uniquely, the one recently reported in \cite{cqg1}. The
spacetime, in general, has non-vanishing energy flux and shear.
The spacetime admits matter with {\it any\/} equation of state.
\end{abstract}

\vspace{1cm} \centerline{Submitted to: Physical Review Letters}

\vspace{1cm}

A self-similar spacetime is characterized by the existence of a
Homothetic Killing vector field \cite{cahilltaub}. For such a
spacetime all points along the integral curves of the Homothetic
Killing vector field are similar. Any spherically symmetric
spacetime is self-similar if it admits a radial area coordinate
$r$ and an orthogonal time coordinate $\tau$ such that for the
metric components $g_{\tau\tau}$ and $g_{rr}$ the following holds
\beq g_{\tau\tau}\left(\,\kappa\tau,\,\kappa r\,\right) &=&
g_{\tau\tau}\left(\,\tau,\,r\,\right) \\
g_{rr}\left(\,\kappa\tau,\,\kappa r\,\right) &=&
g_{rr}\left(\,\tau,\,r\,\right) \eeq for all constants $\kappa>0$.
Consequently, for a self-similar spacetime, the Einstein field
equations reduce to ordinary differential equations.

The main application of  self-similar spacetimes has been in the
cosmological context \cite{cahilltaub,psj,cosmoself}. Such
spacetimes can also describe the gravitational collapse if
suitable matching can be done with an appropriate exterior
spacetime \cite{oripiran,cqg1}. While self-similarity is a strong
restriction of a geometric nature on the spacetime, it has been
successfully exploited in various physical scenarios. (See
\cite{coleycqg} for a recent survey on the importance of
self-similarity in General Relativity.)

A Conformal Killing Vector (CKV) ${\bf X}$ satisfies \be {\cal
L}_{\bf X}\,g_{ij}\;=\;2\,\Phi(x^k)\,g_{ij} \label{ckveq} \ee
where $\Phi(x^k)$ is the conformal factor and $g_{ij}$ is the
spacetime metric tensor. In the case of Homothetic Killing
vectors, $\Phi$ must be a constant. Such vectors scale distances
by the same constant factor and also preserve the null geodesic
affine parameters.

In what follows, we will demand that a spherically symmetric
spacetime admits a Homothetic Killing vector of the form \be
X^a\;=\;(0,f(r,t),0,0) \label{vec1} \ee Typically, the homothetic
Killing vector in suitable coordinates is taken to be \cite{psj}:
\be \bar{X}^a\;=\;(t,r,0,0) \label{vec2} \ee However, any vector
of the form (\ref{vec1}) can be transformed into (\ref{vec2}) via
a coordinate transformation of the form \be R = l(t)
\exp\left(\int f^{-1} d r\right) \qquad T = k(t) \exp\left(\int
f^{-1} d r\right) \label{sstrans}\ee and so we will not loose any
generality\footnote{Note: The general metric admitting
(\ref{vec2}) has metric functions which are functions of $t/r$. If
we invoke (\ref{sstrans}), the resulting metric will not be
diagonal.  The imposition of diagonality of the metric will
require a relationship between $l(t)$ and $k(t)$. Such a relation
can always be imposed.} by concentrating on (\ref{vec1}).

\goodbreak If we require that the general spherically symmetric
line element \be ds^2 \,=\, - e^{2\nu(t,r)} dt^2 \,+\,
e^{2\lambda(t,r)} dr^2 \,+\, Y^{2}(t,r)(d\theta^2 + \sin^2 \theta
d\phi^2) \label{met1} \ee admits (\ref{vec1}), the expression
(\ref{ckveq}) reduces to the system of four equations: \beq
f(r,t)\, \frac{\partial\,\nu}{\partial r}&=& \Phi \label{phinu} \\
\frac{\partial\,f(r,t)}{\partial t} &=& 0 \label{fcr}\\
\left( \frac{1}{Y}\,\frac{\partial\,Y}{\partial r}\,-\,
\frac{\partial\,\nu}{\partial r} \right)\, f(r,t) &=& 0 \\
\left( \frac{\partial\,\lambda}{\partial r}\,-\,
\frac{\partial\,\nu}{\partial r} \right)\, f(r,t)\;+\;
\frac{\partial\,f(r,t)}{\partial r} &=&0 \eeq where $\Phi$ is a
constant. (See \cite{selvan} for a comprehensive description of
the general conformal geometry of (\ref{met1}).)

\goodbreak \noindent Solving the above system of equations, we obtain \beq f &=& F(r)
\label{feqn} \\
Y &=& \tilde{g}(t)\,\exp\left(\int\frac{\Phi}{F(r)}d r\right)
\\ \lambda &=& \int\frac{\Phi}{F(r)}d r \;-\;\log F(r) \;+\; \tilde{h}(t) \\
\nu &=& \int\frac{\Phi}{F(r)}d r + \tilde{k}(t)  \eeq so that the
 spacetime metric becomes {\it separable\/} and is given by \be ds^2\;=\;k^2(t)
\exp\left(2\int\frac{\Phi}{F(r)}d
r\right)\left[-dt^2\;+\;\frac{h^2(t)}{F^2(r)}\,dr^2
\;+\;g^2(t)\left( \,d\theta^2\;+\;\sin^2{\theta}\,d\phi^2
\right)\right] \label{finalmet} \ee

One would now substitute this metric into the Einstein Field
Equations in order to determine the explicit forms of the metric
functions for the choice of the energy-momentum tensor for the
matter in the spacetime. However, this metric has already been
obtained in \cite{cqg1} as:\be
ds^2\;=\;-\,y^2(r)\,A^2(t)\,dt^2\;+\;\gamma^2\,B^2(t)\,\left(\frac{dy}{dr}\right)
^2\,dr^2\;+\;y^2(r)\,R^2(t) \left[ \,
d\theta^2\;+\;\sin^2{\theta}\,d\phi^2 \,\right] \label{met}\ee
where $y(r)$ is an arbitrary function of $r$.

(Using the differential equations solver in {\tt PROGRAM LIE}
\cite{head}, it is easy to verify that the Homothetic Killing
vector \be X = \frac{y}{\Phi\,y'} \frac{\partial\ }{\partial r}
\label{methkv} \ee is admitted by (\ref{met}). See also
\cite{kill}.) The two metrics (\ref{finalmet}) and (\ref{met}) are
clearly identical.

We note that the requirement of the self-similarity of the
spherically symmetric spacetime {\em uniquely\/} fixes the metric
to (\ref{met}).  All metrics which admit a Homothetic Killing
vector will be contained in this metric (or can be transformed to
this form). The spacetime of metric (\ref{met}) admits any
equation of state for the matter in the spacetime. This spacetime
was obtained by first imposing the separability of the metric
functions and then imposing the field equations. {\it It now
follows that the assumption of separability of the metric
functions for a spherically symmetric spacetime and the
satisfaction of the field equations provide us with a self-similar
spacetime and vice-versa.}

The spacetime (\ref{met}) is radiating and shearing. All
spacetimes admitting a Homothetic Killing vector, that have
vanishing or non-vanishing shear and/or heat flux, {\it must} be
contained with this spacetime.  The crucial point here is that, in
those cases, more field equations need to be solved which further
constrains the metric functions. However, the equivalence of the
spacetimes only holds if (\ref{sstrans}) is non-singular. For
example, in the case of the Robertson-Walker spacetime, the
transformation is singular as the function $F(r)\,=\,0$.

\section*{Acknowledgements}

KSG thanks the University of Natal and the National Research
Foundation for ongoing support.  He also thanks CIRI for their
kind hospitality during the course of this work.

\end{document}